\documentclass[aps,prl,twocolumn,preprintnumbers,superscriptaddress,amsmath,longbibliography]{revtex4-1}
\pdfoutput=1
\usepackage{amssymb}
\usepackage{txfonts}
\usepackage{graphicx}
\usepackage{subfigure}
\usepackage{dcolumn}
\usepackage{bm}
\usepackage{color}
\usepackage{upgreek}
\usepackage{physics}

\usepackage{natbib}
\usepackage[colorlinks=true,linkcolor=blue,citecolor=magenta,urlcolor=blue]{hyperref}

\newcommand{\bracket}[3]{\langle#1|#2|#3\rangle}

\begin{document}

\title{Higher-dimensional symmetric informationally complete measurement via programmable photonic integrated optics}

\author{Lan-Tian Feng}
\email{These three authors contributed equally to this work.}
\author{Xiao-Min Hu}
\email{These three authors contributed equally to this work.}
\affiliation
{CAS Key Laboratory of Quantum Information, University of Science and Technology of China, Hefei 230026, China.}
\affiliation{CAS Synergetic Innovation Center of Quantum Information $\&$ Quantum Physics, University of Science and Technology of China, Hefei 230026, China.}
\affiliation{Hefei National Laboratory, University of Science and Technology of China, Hefei 230088, China.}
\author{Ming Zhang}
\email{These three authors contributed equally to this work.}
\affiliation{State Key Laboratory for Modern Optical Instrumentation, Centre for Optical and Electromagnetic Research,
Zhejiang Provincial Key Laboratory for Sensing Technologies, Zhejiang University, Zijingang Campus, Hangzhou
310058, China}
\author{Yu-Jie Cheng}
\author{Chao Zhang}
\author{Yu Guo}
\affiliation{CAS Key Laboratory of Quantum Information, University of Science and Technology of China, Hefei 230026, China.}
\affiliation{CAS Synergetic Innovation Center of Quantum Information $\&$ Quantum Physics, University of Science and Technology of China, Hefei 230026, China.}
\author{Yu-Yang Ding}
\affiliation{Hefei Guizhen Chip Technologies Co., Ltd., Hefei 230000, China.}
\author{Zhibo Hou}
\author{Fang-Wen Sun}
\author{Guang-Can Guo}
\affiliation
{CAS Key Laboratory of Quantum Information, University of Science and Technology of China, Hefei 230026, China.}
\affiliation{CAS Synergetic Innovation Center of Quantum Information $\&$ Quantum Physics, University of Science and Technology of China, Hefei 230026, China.}
\affiliation{Hefei National Laboratory, University of Science and Technology of China, Hefei 230088, China.}

\author{Dao-Xin Dai}
\email{dxdai@zju.edu.cn}
\affiliation{State Key Laboratory for Modern Optical Instrumentation, Centre for Optical and Electromagnetic Research,
Zhejiang Provincial Key Laboratory for Sensing Technologies, Zhejiang University, Zijingang Campus, Hangzhou
310058, China}

\author{Armin Tavakoli}
\email{armin.tavakoli@teorfys.lu.se}
\affiliation
{Physics Department, Lund University, Box 118, 22100 Lund, Sweden.}

\author{Xi-Feng Ren}
\email{renxf@ustc.edu.cn}
\author{Bi-Heng Liu}
\email{bhliu@ustc.edu.cn}
\affiliation
{CAS Key Laboratory of Quantum Information, University of Science and Technology of China, Hefei 230026, China.}
\affiliation{CAS Synergetic Innovation Center of Quantum Information $\&$ Quantum Physics, University of Science and Technology of China, Hefei 230026, China.}
\affiliation{Hefei National Laboratory, University of Science and Technology of China, Hefei 230088, China.}

\begin{abstract}
Symmetric informationally complete measurements are both important building blocks in many quantum information protocols and the seminal example of a generalised, non-orthogonal, quantum measurement. In higher-dimensional systems, these measurements become both increasingly interesting and increasingly complex to implement. Here, we demonstrate an integrated quantum photonic platform to realize such a measurement on three-level quantum systems. The device operates at the high fidelities necessary for verifying a genuine many-outcome quantum measurement, performing near-optimal quantum state discrimination, and beating the projective limit in quantum random number generation. Moreover, it is programmable and can readily implement other quantum measurements at similarly high quality. Our work paves the way for the implementation of sophisticated higher-dimensional quantum measurements that go beyond the traditional orthogonal projections.
\end{abstract}


\maketitle


\section{Introduction}
Standard measurements in quantum theory are projections onto a basis in $d$-dimensional Hilbert space. However, in the modern form of quantum theory, there exist more complex measurements, that can on their own be tomographically complete, with up to $d^2$ outcomes, while remaining  irreducible \cite{DAriano2005}. That is, they cannot be simulated by stochastically combining any other quantum measurements. The most prominent example of such generalised quantum measurements is a symmetric informationally complete (SIC) measurement \cite{Renes2004}. Its defining property, in addition to tomographic completeness and irreducibility, is outcomes being fully  symmetric in Hilbert space. 

Much research has been devoted to theoretically constructing these measurements~\cite{Fuchs2017}. While they find many applications in quantum information, they are also connected to some interpretations of quantum theory \cite{Fuchs2013} and well-known mathematics problems  \cite{Appleby2017}. In quantum information, SIC measurements are natural for many important tasks. Examples are state tomography \cite{Scott2006, Stricker2022}, entanglement detection \cite{Shang2018, Morelli2023} and random number generation \cite{Acin2016, Kaniewski2021}. They are also relevant for quantum certification protocols \cite{Tavakoli2019}, quantum key distribution \cite{Renes2005,Bengtsson2020} and tests of nonlocality and contextuality  \cite{Bengtsson2012, Kaniewski2021}. Naturally, there has been considerable interest in experimentally realising SIC measurements. However, the implementation of generalised measurements comes with additional challenges. It requires control of ancillary quantum systems and entangling gate operations. In the case of SIC measurements, the challenge is particularly pronounced, as one must simultaneously resolve the maximal number of $d^2$ outcomes while also achieving uniform Hilbert space angles by control of populations and phases. Early photonics realisations of SIC measurements are based on simulating it by separately performing several standard measurements and post-selecting on successful outcomes \cite{Medendorp2011, Pimenta2013, Bent2015}. Faithful implementations, using multiple photonic degrees of freedom for system and ancilla, have been achieved for the simplest, qubit, systems, see e.g.~\cite{bian2015realization,zhao2015experimental,hou2018deterministic}. However, it is well-known that already the interferometry required for realising SIC measurements beyond qubit dimensions is much more demanding \cite{Tabia2012}.

Going beyond binary quantum systems is an important frontier in quantum information science. Higher-dimensional systems propel advantages in many communication protocols. In quantum key distribution, higher-dimensional systems can give better key rates and offer important improvements to the tolerance of noise and loss \cite{hu2021pathways, Bulla2023}. Similar high-dimensional advantages are encountered in basic quantum information primitives such as entanglement distribution \cite{Ecker2019,hu2020efficient}. In nonlocality experiments, high-dimensional systems can lead to larger Bell inequality and steering inequality violations \cite{Dada2011, Srivastav2022}. Also in quantum computation, processors have been designed for higher dimensions \cite{Paesani2021, Chi2022}.

\begin{figure*}[t]
\centering
\includegraphics[width=16.0cm]{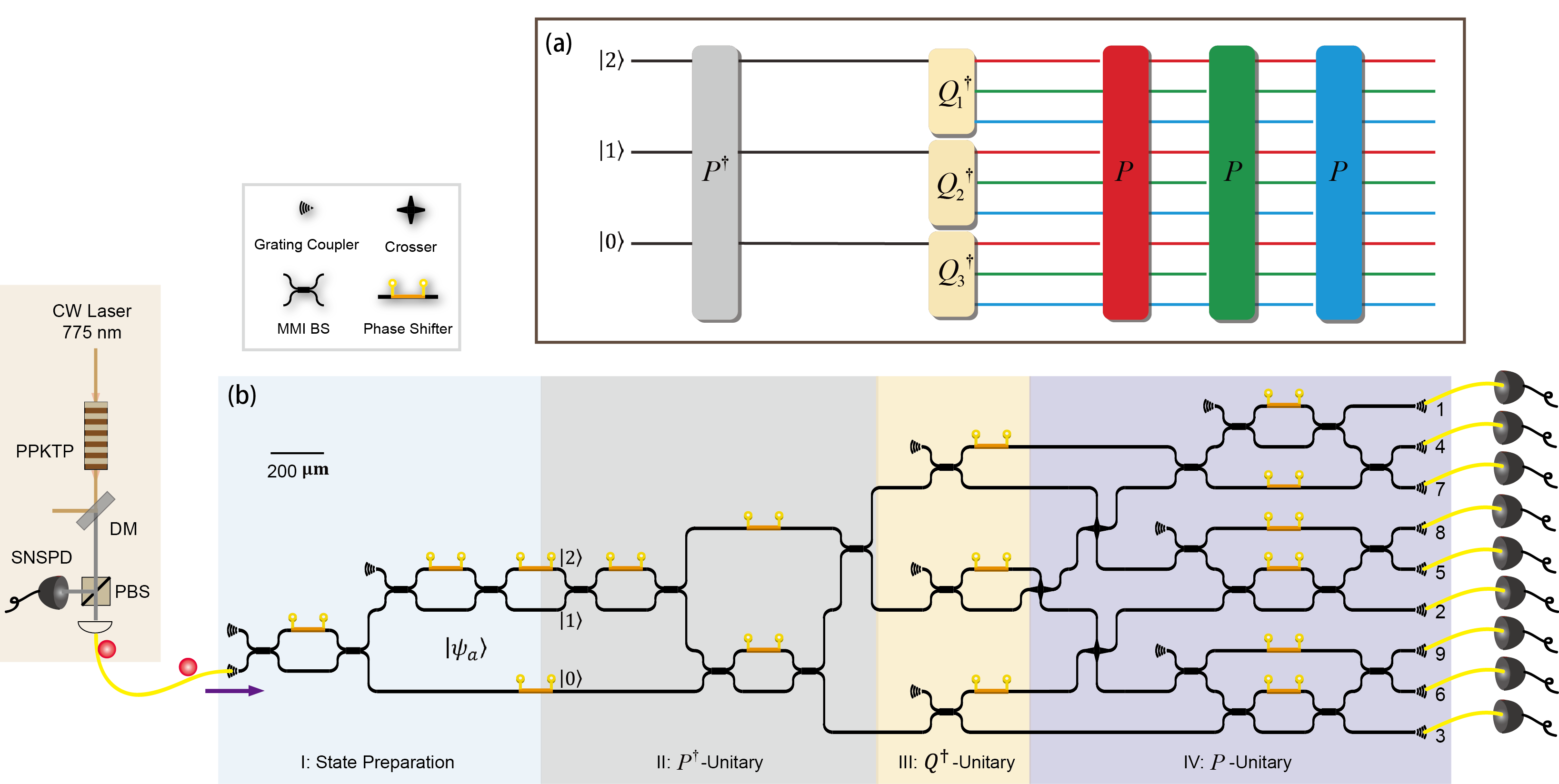}
\caption {Scheme and experimental demonstration. (a) Schematic circuit for implementation of qutrit SIC measurement using multiport devices and qutrit subspace unitaries. The device has three inputs and nine outputs. The three paths on which each of the boxes labeled $P$ operate are colour coded. (b) Experimental demonstration with the silicon quantum photonic processor. I: Prepare qutrit state. The structure constructed by the MZI and phase shifter can achieve any three-dimensional pure-state preparation. II, III, IV: 3-input 9-output measurement, consisting of cascaded three-dimensional subspace unitary operations. II: Three MZIs and three phase shifters achieve the  $P^{{\dag}}$-unitary in modes 1, 2, and 3. III: Perform $Q_1^{{\dag}}$, $Q_2^{\dag}$, and $Q_3^{\dag}$ unitary operations on three paths, respectively, each consisting of one MMI BS and one phase shifter. IV: Three $P$-unitary operations need to be executed in parallel in this region.  In order to facilitate the implementation of three-dimensional subspace unitary operations, we first perform path swapping, swapping paths of the same subspace operation to the same region. Afterward, the $P$-unitary operations are implemented in each subspace. Finally, regions II, III, and IV form a SIC measurement with 3 inputs and 9 outputs. By changing the phase setting of the phase shifter, we can also achieve full outcome mutually unbiased basis measurement. CW Laser: continuous-wave laser; PPKTP: nonlinear crystal; DM: dichroic mirror; PBS: polarizing beam splitter; SNSPD: superconducting nanowire single-photon detector; MMI BS: multimode interference beam splitter.}
\label{fig1}
\end{figure*}

It is a natural endeavor to develop measurement devices for beyond-binary quantum systems, that can faithfully implement not only standard but also generalised quantum measurements. Notably, a higher-dimensional generalised measurement has been reported \cite{Martinez2023}, but it is conceptually different from a SIC measurement because it is neither tomographically complete nor motivated by established quantum information tasks. The biggest challenge in high dimensional SIC measurements is that the complexity of constructing measurements increases rapidly with increasing dimensions. A promising avenue to overcome this challenge is photonic integrated circuits (PICs), which provide high stability and scalability and are developing into a promising platform for quantum photonic experiments \cite{wang2020integrated,Feng2022silicon}. Many quantum applications including logic gates \cite{crespi2011integrated,feng2022transverse}, higher-dimensional entanglement with standard measurements \cite{wang2018multidimensional,lu2020three,Chi2022}, tests of Boson Sampling \cite{crespi2013integrated,arrazola2021quantum}, quantum teleportation \cite{llewellyn2020chip} and frustrated quantum interference \cite{Feng2023chip} have been demonstrated with PICs. Photonic quantum states can be directly generated, manipulated, and detected on integrated chips. Moreover, because of the high interferometric visibility and phase stability, quantum PIC with many passive and active components is particularly attractive to construct large-scale programmable quantum photonic processors. To date, quantum PICs have been expanded to contain thousands of components \cite{bao2023very} and dozens of photons \cite{arrazola2021quantum}. 

Here, we develop a PIC device that can faithfully implement generalised quantum measurements, in the form of SIC measurements on three-level quantum systems. Our device  operates at high quality, clearly achieving the standards necessary both for demonstrating physical properties well beyond those of standard, projective, measurements, and for harvesting the advantages of generalised measurements in quantum information protocols. Specifically, this is demonstrated  through detector tomography, through a range of verification and simulation tasks and through quantum random number generation. In addition, the device is versatile as it can be readily programmed to implement many other measurements. We demonstrate this feature by realising  highly symmetric projections as well as the seminal mutually unbiased bases, and apply them in relevant information tasks.

\section{Scheme for symmetric informationally complete measurement}

A quantum measurement in dimension $d$ with $n$ outcomes is described by a collection of positive semi-definite $d\times d$ matrices $\{E_a\}_{i=1}^n$ such that $\sum_a E_a=\openone$. For irreducible standard measurements, $E_i$ are orthogonal projectors and the number of outcomes is at most $n=d$. For irreducible generalised measurements, one can reach $n=d^2$ outcomes. The SIC measurement corresponds to choosing $E_a=\frac{1}{d}\ketbra{\psi_a}{\psi_a}$ where all distinct pairs of modulus overlaps are identical, namely $|\braket{\psi_j}{\psi_k}|^2=\frac{1}{d+1}$ for every $j\neq k$. A standard instance of a qutrit SIC measurement corresponds to the following nine (unnormalized) vectors $\ket{\psi_a}$:
\begin{equation}\label{SICstate}
\left\{\hspace{2mm}\begin{split}
&(0,1,-1) \qquad (0,1,-\omega) \qquad (0,1,-{\omega}^2)\\&
(-1,0,1) \qquad (-\omega,0,1) \qquad (-{\omega}^2,0,1)\\&
(1,-1,0) \qquad (1,-\omega,0) \qquad (1,-{\omega}^2,0)
\end{split} \hspace{2mm}\right\},
\end{equation}
where $\omega=e^{\frac{2{\pi}i}{3}}$. 

Generalized measurements can always be achieved through standard projection measurements combined with auxiliary systems which can be achieved through a so-called Naimark dilation \cite{peres1990neumark}. When $n>d$ and the measurement operators, $E_a$, are proportional to rank-one projectors onto a vector $e_a$, the $d\times n$ matrix $U=\left(e_1 \ldots e_n\right)$ must be extended to a larger Hilbert space, in which its columns become orthogonal. Hence, we require an $(n-d) \times n$ matrix $W$ such that $U=\left(\begin{array}{c}
V \\
W
\end{array}\right)$ is unitary.  

We consider the Naimark extension for the SIC measurement corresponding to the following unitary matrix,
\begin{equation}
U=\frac{1}{\sqrt{6}}\left(\begin{array}{ccccccccc}
0 & 0 & 0 & -1 & -\omega^2 & -\omega & 1 & \omega & \omega^2 \\
\sqrt{2} & \sqrt{2} & \sqrt{2} & 0 & 0 & 0 & 0 & 0 & 0 \\
1 & \omega^2 & \omega & 1 & \omega & \omega^2 & 0 & 0 & 0 \\
1 & \omega & \omega^2 & 0 & 0 & 0 & -1 & -\omega^2 & -\omega \\
0 & 0 & 0 & \sqrt{2} & \sqrt{2} & \sqrt{2} & 0 & 0 & 0 \\
0 & 0 & 0 & 1 & \omega^2 & \omega & 1 & \omega & \omega^2 \\
-1 & -\omega^2 & -\omega & 1 & \omega & \omega^2 & 0 & 0 & 0 \\
0 & 0 & 0 & 0 & 0 & 0 & \sqrt{2} & \sqrt{2} & \sqrt{2} \\
1 & \omega & \omega^2 & 0 & 0 & 0 & 1 & \omega^2 & \omega
\end{array}\right) .
\end{equation}
Here, the rows 1, 4, and 7 correspond to the matrix $U$ containing the qutrit SIC states \eqref{SICstate}. This matrix is to be applied to an incoming qutrit and initialised ancilla qutrit ($\ket{\psi}\otimes \ket{0}$), with the former ordered on ports  $(1,4,7)$. In order to reduce the complexity of implementing the unitary transformation $U$, we decompose the $9\times 9$ matrix into a sequence of three-dimensional subspace operations. These are shown in Fig. \ref{fig1}a, in terms of unitaries $P $ and $Q_ {1-3}$ (see Supplementary Material), which are unit operations in suitable three-dimensional subspaces \cite{tabia2012experimental}. 

\begin{table*}
\renewcommand\arraystretch{1.3}
\tabcolsep=0.2cm
\begin{tabular}{c|c  c  c  c  c  c  c  c}
\hline \hline Number of outcomes & 2 & 3 & 4 & 5 & 6 & 7 & 8 & 9 \\
\hline$v_{\text {crit }}$ for SIC measurement  & 0.4330 & 0.7931 & 0.8898 & 0.9269 & 0.9571 & 0.9714 & 0.9874 & 1 \\
\hline$v_{\text {crit }}$ for tomographic reconstruction & 0.4549 & 0.8332 & 0.9320 & 0.9690 & 0.9918 & 1 & 1 & 1 \\
\hline Maximal $p_{\text {suc }}$ for exact preparations & 0.2073 & 0.2874 & 0.3088 & 0.3171 & 0.3238 & 0.3270 & 0.3305 & $1 / 3$ \\
\hline Maximal $p_{\text {suc }}$ for imprecise preparation & 0.2169 & 0.3082 & 0.3231 & 0.3277 & 0.3313 & 0.3325 & 0.3333 & $1 / 3$ \\
\hline \hline
\end{tabular}
\caption{Critical measurement visibilities and critical success probabilities in state discrimination for simulation with classical randomness and $n$-outcome quantum measurements.} \label{tab:simulation}
\end{table*}


\section{Implementation}
We use a reconfigurable silicon photonic integrated circuit to implement the high-dimensional SIC measurement scheme. Silicon photonics show sizable advantages in scaling and functionality with low-cost, CMOS-compatible fabrication and ultra-high integration density \cite{thomson2016roadmap}. As shown in Fig. \ref{fig1}b, the device we developed consists of a cascade of Mach-Zehnder interferometers (MZIs) and multiple phase shifters, which are programmed to realize the expected operations given below. The total chip size is just 3.3$\times$1.0$\,\rm{mm}^2$, and all phase shifters are characterized with average interference visibility over 0.99. 

Depending on the function, the device is divided into two parts, the state preparation part (region I) and the measurement part (regions II, III, IV). 
In the state preparation part, any three-dimensional pure state can be constructed using two MZIs and phase shifters. The subsequent measurement apparatus, with three inputs and nine outputs, can be switched between several different settings. For example, by adjusting the phase shifter, one can switch from a nine-outcome SIC measurement to different sets of mutually unbiased basis measurements. In the case of SIC measurement, in our scheme, we only need to implement the unitary operation on three-dimensional subspaces. As shown in Fig~\ref{fig1}, we sequentially perform the unitaries $P^{\dag}$, $Q^{\dag}$, and $P$ operations on the three-dimensional subspaces in the II, III, and IV regions, ultimately resulting in 9 outputs. In contrast, mutually unbiased bases, being standard projective measurements, require only a unitary operation of 3 inputs and 3 outputs. They can be implemented by resetting the phase of the phase shifter. The specific phase shifter settings and further details on the PIC device can be found in the supplementary materials.

In the experiment, a CW laser at 775 nm is used to pump a type II PPKTP crystal to generate correlated photon pairs \cite{feng2022transverse}. They are coupled into different singe-mode fibers, and one photon is detected directly to herald the existence of another. The photons show a coherence time of about 1.5 ps. The heralded single photons are coupled into the silicon quantum photonic processor through the single-mode fiber and on-chip grating coupler, and the qutrit states are generated and analyzed on the single chip.



\section{Tomography and verification of the SIC measurement.}
A central question is to gauge how accurately the measurement implemented on our device corresponds to the SIC measurement. For this purpose, we discuss several different verification methods. Details are given in Supplementary Material.

\subsection*{Tomography}
The most immediate approach is to perform detector tomography for the device. To this end, we have prepared the tomographically complete set of qutrit states $\{\ket{\psi_a}\}_{a=1}^9$ given in Eq.~\eqref{SICstate}. Probabilities were inferred from the relative frequencies in the nine different ports of the device. The tomographic reconstruction of the measurement operators was based on a standard maximum likelihood method \cite{fiuravsek2001maximum}. The estimated fidelity of each of the operations corresponding to the nine possible outcomes are $0.9553\pm0.0034$, $0.9649\pm0.0029$, $0.9703\pm0.0028$, $0.9842\pm0.0023$, $0.9663\pm0.0030$, $0.9554\pm0.0028$, $0.9477\pm0.0024$, $0.9668\pm0.0029$ and $0.9734\pm0.0030$ respectively. The average fidelity is $0.9648\pm0.0010$.

\begin{figure*}[t]
\centering
\includegraphics[width=13.0cm]{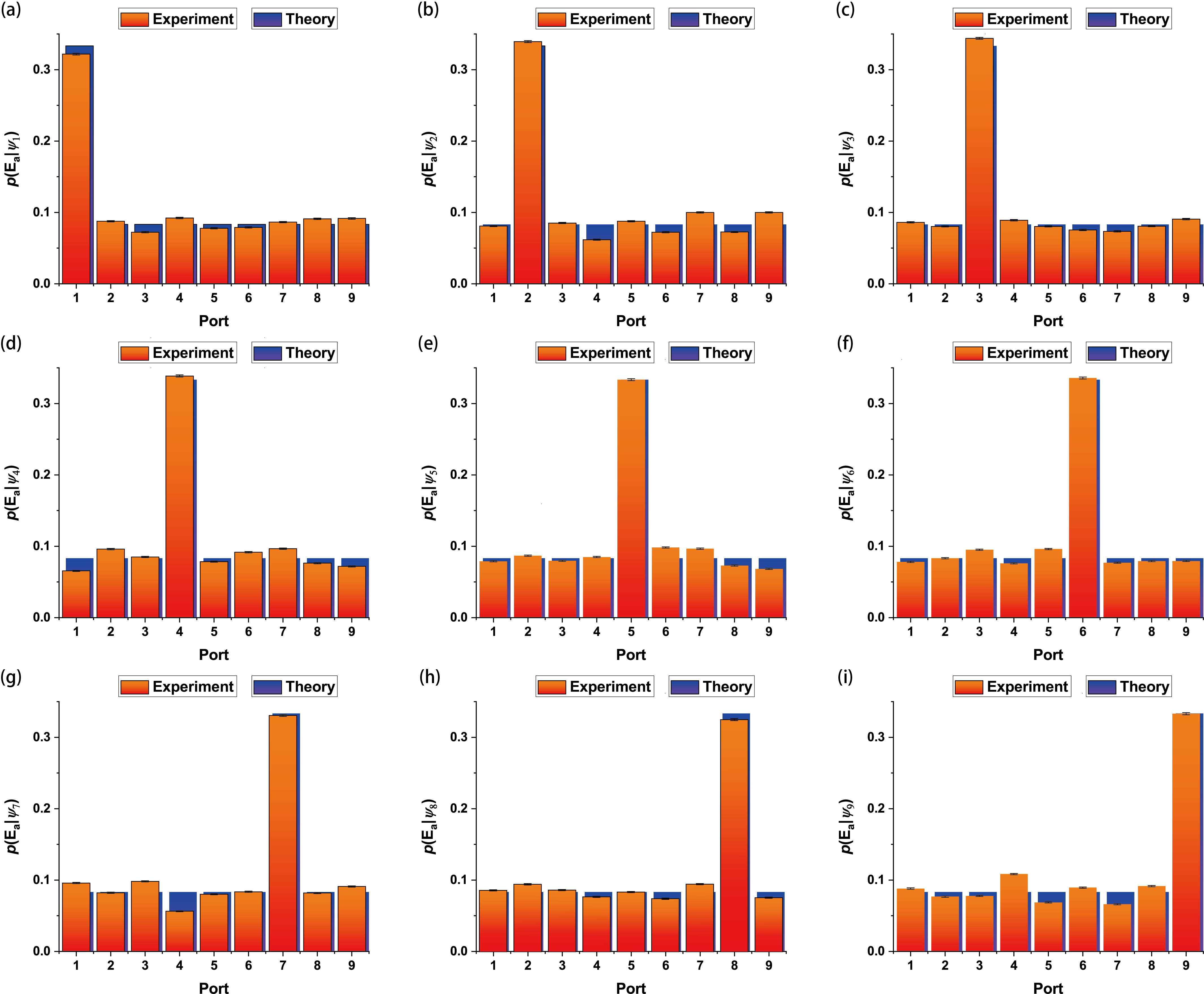}
\caption {Quantum state discrimination results. (a)-(i) represents the probability distribution over the nine outcomes when the input state is $|\psi_1\rangle-|\psi_9\rangle$. The horizontal axis represents the outcome of the SIC measurement and the vertical axis represents the probability of each outcome. The blue column represents the theoretical value. The red column represents the experimental value.}
\label{fig3}
\end{figure*}

\subsection{Simulation of tomographic reconstruction}
To benchmark the quality of our lab measurement, we investigate whether it can be simulated by stochastically combining quantum measurements with at most $n$ outcomes. If no $n$-outcome simulation exists, we conclude that must genuinely necessitate at least $n+1$ outcomes.  Notably, defying a simulation with $n=3$ implies a truly generalised, non-projective, measurement. 

For each $n=2,\ldots,9$, we either construct an explicit simulation or quantify its failure through the minimal rate of noise that must be added to the measurement to make the simulation possible. To this end, we take the artificially mix the lab measurement with noise of uniform spectral density, $E_a^{(v)}=v E_a^{\text{tomo}}+(1-v) \operatorname{Tr}\left(E_a^{\text{tomo}}\right) \frac{\openone}{3}$, 
for some visibility parameter $v\in[0,1]$. Here,  the lab measurement is represented by the tomographic reconstruction. Now,  we seek the largest visibility, $v^{\text{crit}}_n$, for which $\{E_a^{(v)}\}$ admits an $n$-outcome simulation. Notably, any value $v^{\text{crit}}_n<1$ implies that $E_a^{\text{tomo}}$ cannot be simulated. Using semidefinite programming  \cite{tavakoli2023semidefinite}, following the technique of Ref.~\cite{Oszmaniec2017}, we can compute $v^{\text{crit}}$. The results are displayed in Table~\ref{tab:simulation}. The lab measurement clearly defies the projective limit since $v_3^{\text{crit}}=0.8332$, instead requiring  $n=7$ outcomes for a simulation.

However, this procedure has a significant drawback that may lead to an underestimation of the quality of the lab measurement. The tomographic reconstruction depends significantly on both the specific data analysis method and it is vulnerable to tiny experimental deviations in the state preparation \cite{Rosset2012}. These effects are especially relevant in our case, due to the small gap between the noise rates for large $n$ in the first row of Table~\ref{tab:simulation}. Next, we set out to eliminate these issues.


\subsection{Simulation of quantum state discrimination}
We now use the basic quantum information primitive of state discrimination to benchmark our implementation of the SIC measurement. Consider that we draw, with uniform probability, one of the nine states in Eq.~\eqref{SICstate} and try to recover the classical label via a measurement. The success probability of state discrimination is  $p_\text{suc}=\frac{1}{9}\sum_{a=1}^9 \bracket{\psi_a}{E_a}{\psi_a}$.  In particular, the best measurement, namely the SIC measurement, would achieve $p_\text{suc}=\frac{1}{3}$. We have used semidefinite programs to compute the largest value of $p_\text{suc}$ achievable with arbitrary quantum measurements with at most $n$ outcomes. The results are shown in the third row of Tab~\ref{tab:simulation}. In particular, projective measurements satisfy the tight inequality $p_{\text {suc }} \lesssim 0.2874 $. 

The probability distributions in our implementation of the state discrimination are shown in Fig~\ref{fig3}. The results closely agree with the theoretical prediction for the SIC measurement.  We observe $p_{\text{suc}}=0.3335\pm0.0005$ from a total of roughly $9\times 10^{5}$ counts. This is far above the projective measurement limit and, notably, even somewhat about the limit for the SIC measurement. However, the latter is statistically insignificant. Putting the experimental value even five standard deviations below the mean still exceeds the eight-outcome measurement limit  $p_{\text{suc}}=0.3305$. The $p$-value, obtained from the Hoeffding inequality \cite{Hoeffding1963}, for the mean violation of the eight-outcome limit is of order  $10^{-7}$. This implies a  genuine nine-outcome measurement, which is a substantial improvement on the tomography-based approach. Indeed, the tomographic reconstruction would predict a far smaller value ($p_{\text{suc}}\approx 0.3217$) than observed in the lab.

The verification of the measurement can be further strengthened by eliminating the assumption that the nine preparations are exactly controlled. Following the ideas of \cite{Tavakoli2021sdi}, we permit any qutrit state with limited infidelity to the target preparation. Our average measured preparation infidelity is $7.3\times 10^{-3}$. In the final row of Tab~\ref{tab:simulation}, we have estimated the resulting correction of  $p_{suc}$ for $n$-outcome measurements. While if the preparation device optimally leverages the allowed infidelity, it becomes impossible to beat the eight-outcome limit, our reported mean still exceeds the seven-outcome limit.

\section{Protocols}
We consider the use of the measurement device for protocols beyond the purpose of benchmarking the device.  In our three case studies,  we showcase also the programmability of the measurement device. Details on the protocols and our experiments are found in Supplementary Material.

\subsection{Quantum random numbers}

\begin{figure*}[t]
\centering
\includegraphics[width=13.0cm]{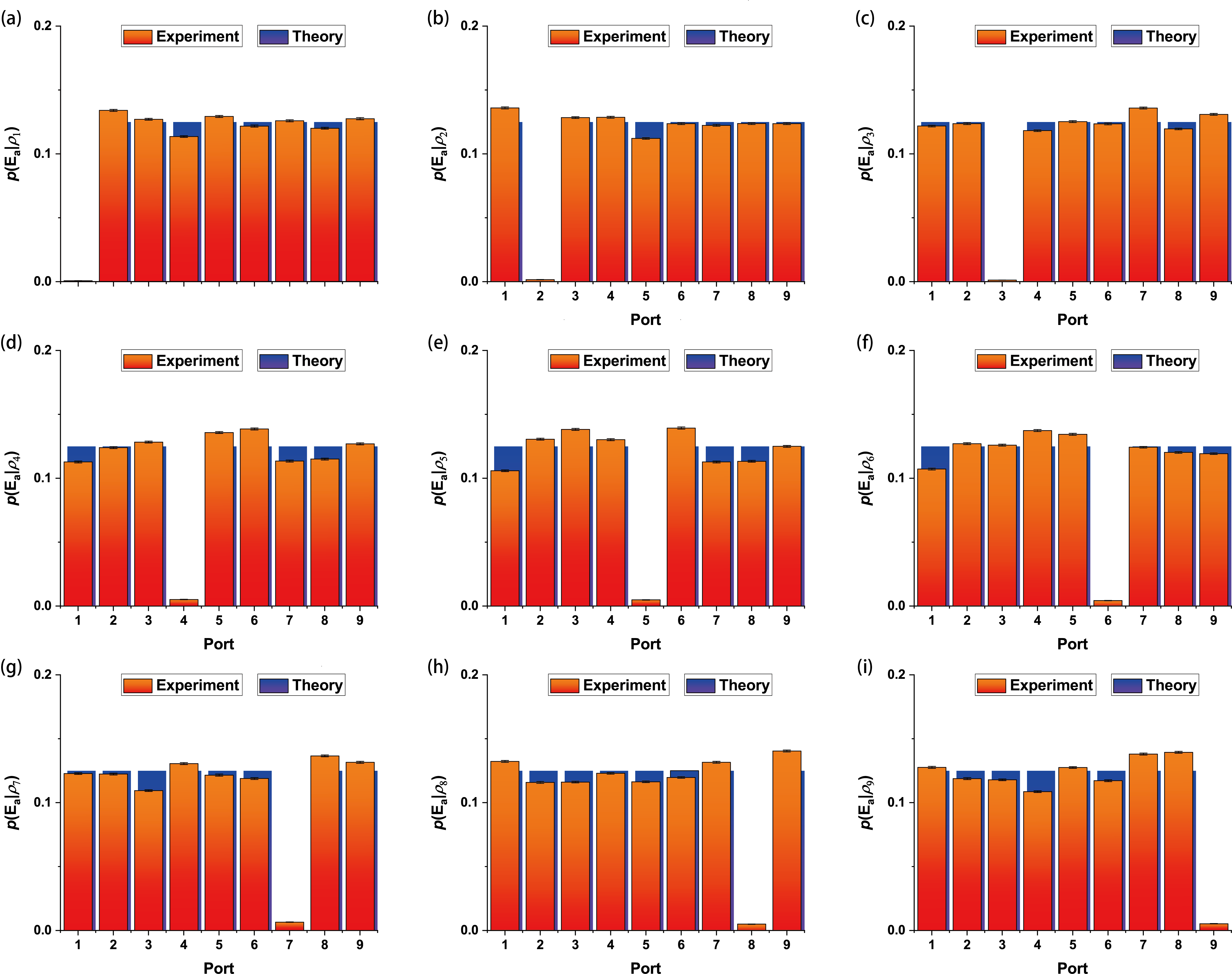}
\caption {Probability distribution for random number generation. (a)-(i) represents the probability distribution over the nine possible outcomes when the input state is $\rho_1-\rho_9$. The horizontal axis represents the outcome of the SIC measurement, while the vertical axis represents the probability of each outcome. The blue column represents the theoretical value, while the red column represents the experimental value. The protocol corresponds to quantum state exclusion.}
\label{fig4}
\end{figure*}
Having no more than $d$ outcomes, irreducible projective measurements cannot generate more than $\log(d)$ bits of randomness. In contrast, generalised measurements can break this limit. The SIC measurement is ideal for the purpose as it features a maximal number of outcomes and is fully unbiased.   In our analysis, we treat the device as a black box and assume that we can probe it with known qutrit states; a so-called measurement-device-independent scenario.  Our protocol is based on quantum state exclusion: we probe the SIC measurement with preparations $\rho_x=\frac{1}{2}\left(\openone-\left|\psi_x\right\rangle\left\langle\psi_x\right|\right)$. Thus, each $\rho_x$ is mixed and orthogonal to the SIC state $\ket{\psi_x}$. The prediction is then a distribution $p(a|x)=\frac{1-\delta_{a,x}}{8}$, which is zero if $a=x$ and uniform otherwise. It has $R=-\log(1/8)=3$ bits of randomness for any choice of $x$.

The probabilities estimated in our implementation are displayed in Fig.~\ref{fig4}. They correspond accurately to the prediction. The randomness of our data can be evaluated using the methods of  \cite{Skrzypczyk2017} but the evaluation ultimately depends on how relative frequencies are chosen to correspond to probabilities. While this can make the final randomness vary to a small extent, it is consistently well above the projective limit of $R=\log(3)\approx 1.58$ bits. For instance, if we permit five (twelve) standard deviations in fluctuations about the relative frequency for $a=x$ ($a\neq x$), we obtain  $R=1.72$ bits. Roughly the same is found when permitting fluctuations corresponding to ten percent of each relative frequency. A small increase is found when the randomness is calculated from the tomographic reconstruction, where fluctuations are not taken into account.

\subsection{Beating dense coding protocols with single qutrits}
Next, we program our device to perform a complete set of four mutually unbiased basis projections. We use this to produce qutrit correlations that elude the typically stronger resource of qubit communication assisted by qubit entanglement, i.e.~protocols \`a la dense coding \cite{Pauwels2022}. We measure each of the nine SIC states \eqref{SICstate} in each of the four bases, with the goal of avoiding a forbidden outcome in each setting. This protocol is rather versatile; it has been used for quantum games  \cite{Emeriau2022}, proofs of contextuality  \cite{Bengtsson2012} and entanglement detection schemes  \cite{Huang2021}. The average probability, $S$, of observing the forbidden event is zero in a perfect implementation but cannot be reduced further than $S_{\text{DC}}\approx 8.0 \times10^{-3}$ in any dense coding protocol. In our implementation, we collect roughly $4.5\times 10^{4}$ counts for each of the 36 settings of the experiment. We observe  $S_{\text{exp}}=(2.39\pm0.04)\times 10^{-3}$, which clearly demonstrates a qutrit advantage.


\subsection{Randomised matching}
Finally, we program the device to implement a set of four rank-one projections that together form a maximally distinguishable set of fully symmetric vectors. These are states $\ket{\phi_a}$ with the property that $|\braket{\phi_a}{\phi_{a'}}|^2$ is constant and minimal for all $a\neq a'$. For $a=0,1,2,3$, they are given by $\ket{\phi_a}=\frac{1}{\sqrt{3}}\left(1,i^a,(-1)^a\right)^T$. Such projections have useful e.g.~for entanglement detection  \cite{Morelli2023, shi2023entanglement} and uncertainty relations \cite{Rastegin2023}. We use them to beat the classical limit in a simple communication problem. 

Alice and Bob privately and randomly select four-valued symbols $x$ and $y$ and attempt to decide whether their selections are identical, but may only communicate a trit message. If identical, Bob must output $a=1$ and if not he must output $a=0$. The average success probability of correctly answering the random matching question is given by $T=\frac{1}{16} \sum_{x, y=1}^4 p\left(a=\delta_{x, y}| x, y\right)$. It can be shown that classical protocols cannot exceed $T_{\text{C}}=\frac{7}{8}$ but a qutrit-based protocol, where Bob's affirmative outcome ($a=1$) corresponds to a projection onto $\ket{\phi_a}$, achieves  $T_{\text{Q}}=\frac{11}{12}\approx 0.9167$. 

In our implementation, we have achieved $T_{\text{exp}}=0.9144 \pm0.0003$, which significantly outperforms the classical limit and nearly reaches the quantum limit.  The statistic corresponds to a total of roughly $8.2\times 10^5$ counts. 


\section{Conclusion}
We have reported on a photonic integrated circuit for implementing a three-dimensional symmetric informationally complete measurement. This demonstrates a faithful realisation, beyond qubit systems, of the perhaps most celebrated type of generalised quantum measurement. Importantly, the quality of the implemented measurement is showcased in a variety of tasks, where it significantly exceeds the limitations of standard, projective, measurements. In principle, our work can be extended to generalized measurements of any dimension combined with mature integrated optics. The results showcase the potential of using integrated photonics towards implementing sophisticated quantum measurements while maintaining both the high quality necessary for harvesting their advantages and the scalability necessary to access such operations in truly higher-dimensional systems. The scalability and programmability of integrated optics for high-dimensional generalized measurement can also be combined with existing light source preparation, transmission, and operation technologies to achieve quantum information tasks at high  performance standards.





\textbf{Funding.}

This work was supported by the Innovation Program for Quantum Science and Technology (Nos. 2021ZD0301500, 2021ZD0303200, 2021ZD0301200), the National Natural Science Foundation of China (NSFC) (Nos. T2325022, 62275240, 62061160487, 12004373, and 12374338), Key Research and Development Program of Anhui Province (2022b1302007), CAS Project for Young Scientists in Basic Research (Grant No. YSBR-049), USTC Tang Scholarship, Science and Technological Fund of Anhui Province for Outstanding Youth (2008085J02), the Postdoctoral Science Foundation of China (No. 2021T140647) and the Fundamental Research Funds for the Central Universities.   A.T. is supported by the Wenner-Gren Foundation and by the Knut and Alice Wallenberg Foundation through the Wallenberg Center for Quantum Technology (WACQT).

\textbf{Acknowledgment.} 

This work was partially carried out at the USTC Center for Micro and Nanoscale Research and Fabrication.

\textbf{Disclosures.}

The authors declare no conflicts of interest.

\textbf{Data availability.} 

Data underlying the results presented in this paper are not publicly available at this time but may be obtained from the authors upon reasonable request.

\textbf{Supplemental document.} 

See Supplement 1 for supporting content.
\bibliography{references.bib}

\end{document}




\section{More details about chip characterization and experiment setup}
\begin{figure}[h]
\centering
\includegraphics[width=15.0cm]{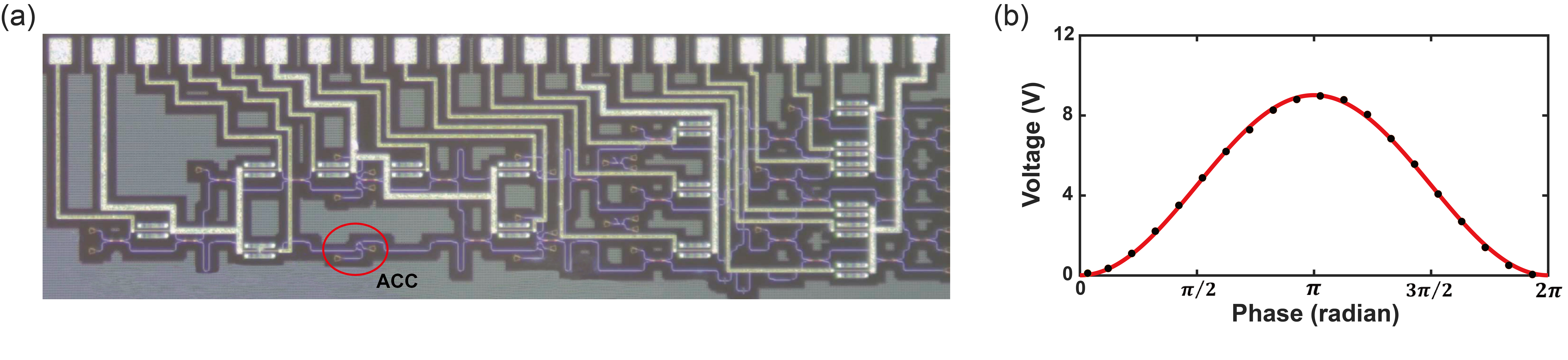}
\caption{\textbf{Experimental setup.} (a) Micrograph of the integrated silicon chip, and the chip size is 3.3$\times$1.0$\,\rm{mm}^2$. ACC: auxiliary calibration coupler. (b) One typical interference fringe of the phase shifter. Visibility of the fringe is 99.8\%.}
\label{supfig1}
\end{figure}

The device we developed was fabricated in a standard silicon photonics foundry. It comprises 29 grating couplers, 9 crossers, 21 multimode interference beam splitters, and 16 phase shifters. All on-chip phase shifters utilize thermo-optical effects to change the phases with millisecond response levels, and they have a similar resistance value of around 500 Ohm. Due to the existence of cascaded phase shifters in the scheme, we set extra auxiliary couplers to facilitate the phase calibration, as shown in Fig.~\ref{supfig1}a. In the experiment, the chip is fixed to one aluminum plate to dissipate the heat generated by the chip. The corresponding phase to the loaded electrical power of the phase shifter can be regarded as a cosine function, as one example shown in Fig.~\ref{supfig1}b. The interference fringe is fitted with $\frac{1}{2}-\frac{V}{2} \cos(\varphi)$, where $V$ is the fringe visibility, $\varphi$ is the phase in the interferometer. The fringe visibility $V$ is defined as $V=(d_{\rm{max}}-d_{\rm{min}})/(d_{\rm{max}}+d_{\rm{min}})$, where $d_{\rm{max}}$ and $d_{\rm{min}}$ are the maximum and minimum of the fitted data, respectively. We calibrated all phase shifters one by one, and they can achieve 2$\pi$ phase adjustment with power consumption lower than 27$\,$mW. In addition, the obtained average interference visibility was over 0.99. The total loss of the chip is 12$\,$dB, including 5$\,$dB per grating coupler, 0.4$\,$dB per auxiliary calibration coupler, and 1.2$\,$dB transmission loss.

In the experiment, we use a type II PPKTP crystal via spontaneous parametric down-conversion processes to generate correlated photon pairs. The crystal is 1$\,$cm long with a cross-section of 1$\times$2$\,\rm{mm}^2$ and poling period of 46.2$\,{\upmu}$m. Since the photon pairs are in opposite polarization, we use a polarization beam splitter to separate them and couple them into different singe-mode fibers. Photons are recorded by two superconducting nanowire single-photon detectors (SCONTEL, the dark count rate is 100 Hz, the detection efficiency is 80\%). With an 80~mW pump, the source's brightness is detected at about 160$\,$kHz. In the chip test, one photon is detected directly and the heralded single photons are input into the chip through the single-mode fiber and on-chip grating coupler. A fiber polarization controller is used to ensure maximum efficiency. The fiber alignment is optimized using a four-dimensional displacement table, and the coupling angle is $12^{\circ}$. The electrical signals generated by detectors are analyzed through a time-correlated single photon counting system with a coincidence window of 0.8 ns. 

\section{Subspace decomposition of unitary matrix $U$}

The circuit for implementing the $9\times 9$ unitary $U$ can be simplified into a sequence of unitary qutrit subspace operations \cite{tabia2012experimental}. After simplification, only $P$ and $Q_{1-3}$ unitary operations in suitable three-dimensional subspaces need to be implemented to realise $U$. The subspace unitaries are 
\begin{equation}
P=\frac{1}{\sqrt{3}}\left(\begin{array}{ccc}
1 & 1 & 1 \\
\omega^2 & \omega & 1 \\
\omega & \omega^2 & 1
\end{array}\right),
\end{equation}

\begin{equation}
Q_1=\frac{1}{\sqrt{6}}\left(\begin{array}{ccc}
0 & i\sqrt{3} & -i\sqrt{3} \\
\sqrt{2} & \sqrt{2} & \sqrt{2} \\
2 & -1 & -1
\end{array}\right),
\end{equation}

\begin{equation}
Q_2=\frac{1}{\sqrt{6}}\left(\begin{array}{lcc}
-i\sqrt{3} & 0 & i\sqrt{3} \\
\sqrt{2} & \sqrt{2} & \sqrt{2} \\
-\omega^2 & 2 \omega^2 & -\omega^2
\end{array}\right),
\end{equation}

\begin{equation}
Q_3=\frac{1}{\sqrt{6}}\left(\begin{array}{lcc}
i\sqrt{3} & -i\sqrt{3} & 0 \\
\sqrt{2} & \sqrt{2} & \sqrt{2} \\
-\omega & -\omega & 2 \omega
\end{array}\right) ,
\end{equation}
where $\omega=e^{\frac{2{\pi}i}{3}}$.

\begin{figure}[h]
\centering
\includegraphics[width=14.0cm]{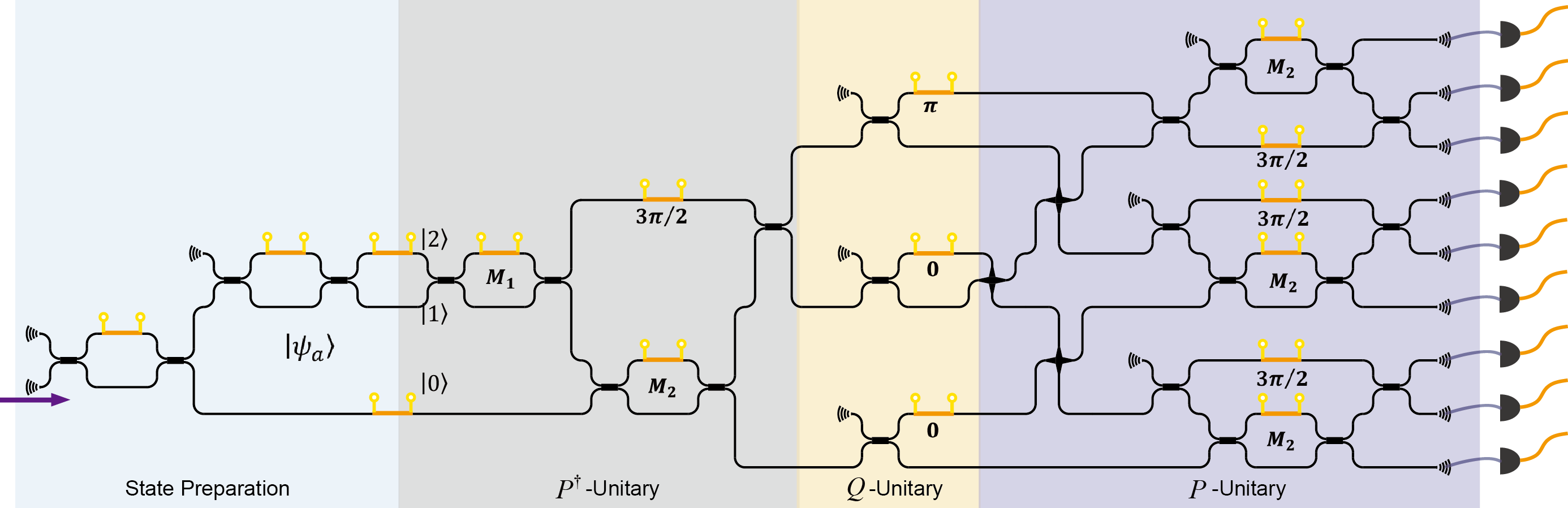}
\caption{\textbf{Phase shifter setting for SIC measurement.} Each MZI is set to realize the indicated unitary operation $M_1=\frac{1}{\sqrt{2}}\left(\begin{array}{cc}
1 & 1 \\
1 & -1\end{array}\right)$ or $M_2=\frac{1}{\sqrt{3}}\left(\begin{array}{cc}
1 & \sqrt{2} \\
\sqrt{2} & -1\end{array}\right)$. } 
\label{supfig2}
\end{figure}

The specific phase shifter setting for the symmetric informationally complete (SIC) measurement is shown in Fig.~\ref{supfig2}. 


\section {Preparation of mixed state $\rho_1$-$\rho_9$} 

In measurement-device-independent randomness generation, noisy states are prepared by randomly switching between two orthogonal states of each SIC projector. The selected unnormalized orthogonal states for each SIC-projector are $|\psi_1\rangle:(1,0,0),(0,1,1),|\psi_2\rangle:(1,0,0),(0,1,\omega),|\psi_3\rangle:(1,0,0),(0,1,{\omega}^2),|\psi_4\rangle:(0,1,0),(1,0,1),|\psi_5\rangle:(0,1,0),(\omega,0,1),|\psi_6\rangle:(0,1,0),({\omega}^2,0,1),|\psi_7\rangle:(0,0,1),(1,1,0),|\psi_8\rangle:(0,0,1),(1,\omega,0),|\psi_9\rangle:(0,0,1),(1,{\omega}^2,0)$. In the test, the switch period is 100 ms and the integral time for each point is chosen as 20$\,$s.

\section{Quantum state and measurement tomography}

\begin{figure}[h]
\centering
\includegraphics[width=15.0cm]{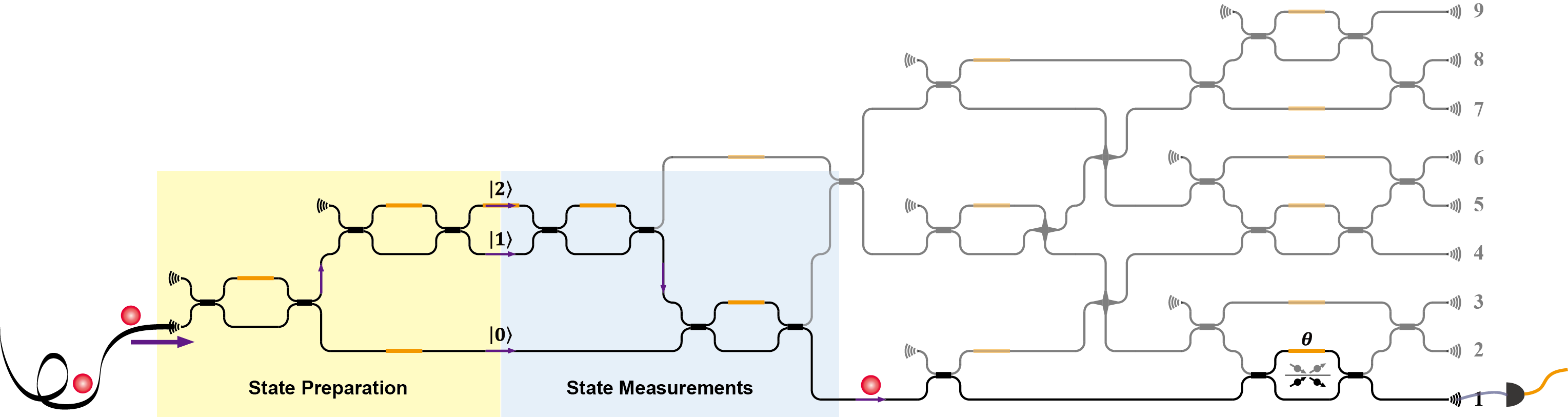}
\caption{\textbf{Experiment setting for quantum state tomography.} The state measurement process is treated as the inverse process of the state preparation. The black line routes the expected measurements to the output port. Phase $\theta$ is fixed to maintain the interferometer fully reflected (shown in the inset).}
\label{supfig3}
\end{figure}

We use the quantum state tomography technique to evaluate the differences between prepared states in the test and expected ideal quantum states $|\psi_x\rangle$. It can reconstruct the density matrix of the system through a series of complementary measurements on a large ensemble of identically prepared copies. For qutrit system, a set of measurements constructed from basis states, $|j\rangle$, and two-state superpositions, $|p\rangle$ and $|q\rangle$, is chosen, where $|p\rangle=(|j\rangle+|k\rangle)/\sqrt{2}$, $|q\rangle=(|j\rangle+i|k\rangle)/\sqrt{2}$, and $j,k\in\{0,1,2\}$. The experiment setting for quantum state tomography is shown in Fig.~\ref{supfig3}, and qutrit states are generated and analyzed on one single chip. The detector is fixed to one output port (Port 1), and the \textit{State Measurements} region can route any measurements in set to the output port. To obtain a physical density matrix, the maximum-likelihood-estimation (MLE) method is used to reconstruct the state density matrix. 

The fidelity between the reconstructed density matrix $\hat{\rho}_{\rm{mea}}$ and the ideal one $\hat{\rho}_{\rm{ideal}}$, defined as $F=\rm{Tr}(\hat{\rho}_{\rm{mea}}\hat{\rho}_{\rm{ideal}})$, where $\hat\rho_{\rm{ideal}}=|\psi_x\rangle\langle\psi_x|$, characterizes the quality of the prepared state. For the ideal state preparation process, $F=1$. The raw fidelities of the prepared nine qutrit quantum states are estimated to be $F_{|\psi_1\rangle}=0.9916\pm0.0003$, $F_{|\psi_2\rangle}=0.9980\pm0.0003$, $F_{|\psi_3\rangle}=0.9890\pm0.0020$, $F_{|\psi_4\rangle}=0.9920\pm0.0004$, $F_{|\psi_5\rangle}=0.9939\pm0.0017$, $F_{|\psi_6\rangle}=0.9885\pm0.0008$, $F_{|\psi_7\rangle}=0.9942\pm0.0003$, $F_{|\psi_8\rangle}=0.9911\pm0.0023$ and $F_{|\psi_9\rangle}=0.9960\pm0.0003$. 
The averaged fidelity is $\bar{F}=0.9927\pm0.0009$, confirming the near-perfect state preparation process. 

We did measurement tomography following the method of Ref.~\cite{fiuravsek2001maximum}. The probability $p_{lm}$ that the apparatus will respond with the positive operator-values measure (POVM) $\Pi_{l}$ when measuring the quantum state can be expressed as $p_{lm}={\rm Tr}[\Pi_{l}\rho_{m}]$. Assuming that the theoretical detection probability $p_{lm}$ can be replaced with relative experimental frequency
$f_{lm}$, we may write ${\rm Tr}[\Pi_{l}\rho_{m}]=\sum_{i,j=1}^{N}\Pi_{l,ij}\rho_{m,ji}=f_{lm}$, 
where $N$ is the dimension of Hilbert space on which the operators $\Pi_{l}$ act. 
The density matrix $\rho_{m}$ chosen in this process is the same as projection operators $\left|\psi_{j}\right\rangle\left\langle \psi_{j}\right|$ in state tomography. The MLE method is also used in this process to ensure $\Pi_{l}\geq0$ and $\sum_{l=1}^{k}\Pi_{l}=I$, 
where $k$ is the number of measurement outcomes.  

Consider two measurements $\left\{E_j\right\}_{j=1}^M$ and $\left\{E'_{j}\right\}_{j=1}^M$ on a
$d$-dimensional Hilbert space. Associate two quantum states: $\sigma=\frac{1}{d} \sum_{j=1}^M E_j \otimes |j\rangle\langle j|$ and $\sigma^{\prime}=\frac{1}{d} \sum_{j=1}^M E'_{j} \otimes|j\rangle\langle j|$, where $|j\rangle$ form an orthonormal basis for an ancilla system. The fidelity between the two measurements is defined as the fidelity between the two \cite{hou2018deterministic}, namely
\begin{equation}
    F\left(\sigma, \sigma^{\prime}\right):=\left(\operatorname{tr} \sqrt{\sqrt{\sigma} \sigma^{\prime} \sqrt{\sigma}}\right)^2=\left(\sum_{j=1}^M w_j \sqrt{F_j}\right)^2,
\end{equation}
where $w_j=\frac{\sqrt{\operatorname{tr}\left(E_j\right) \operatorname{tr}\left(E_{j}^{\prime}\right)}}{d}$, and $F_j=F\left(\frac{E_j}{\operatorname{tr}\left(E_j\right)}, \frac{E_{j}^{\prime}}{\operatorname{tr}\left(E_{j}^{\prime}\right)}\right)$ is the fidelity between the two
normalized POVM elements $\frac{E_j}{\operatorname{tr}\left(E_j\right)}$ and $\frac{E_{j}^{\prime}}{\operatorname{tr}\left(E_{j}^{\prime}\right)}$.


The estimated fidelity of each of operations corresponding to the nine possible outcomes are $0.9553\pm0.0034$, $0.9649\pm0.0029$, $0.9703\pm0.0028$, $0.9842\pm0.0023$,$0.9663\pm0.0030$, $0.9554\pm0.0028$,$0.9477\pm0.0024$,$0.9668\pm0.0029$ and $0.9734\pm0.0030$, respectively. The average fidelity is $0.9648\pm0.0010$.





\section{Simulation with fewer-outcome measurements}

\subsection*{Computing criticial visibility}

Consider a $d$-dimensional quantum measurement $\{E_a\}$ with $a=1,\ldots,m$ outcomes. To this measurement, we apply depolarisation noise, namely we construct a new measurement parameterised by a visibility $v\in[0,1]$ as $E_a^{(v)}=v E_a+(1-v)\Tr(E_a)\frac{\mathbbm{1}}{d}$. We aim to simulate this measurement by using classical randomness and other, arbitrary, quantum measurements that have no more than $n$ outcomes. There are $\binom{m}{n}$ ways of selecting the non-vanishing outcomes in such a quantum measurement. We list the non-vanishing outcomes in a tuple, corresponding to some $n$-cardinality subset of $\{1,\ldots,m\}$, and let  $S_n$ denote the set of all such tuples. We can then select  $T\in S_n$ with probability $p(T)$ and assign a corresponding quantum measurement $\{M_{a|T}\}$ with the constraint that $M_{a|T}=0$ if $a\notin T$. For simplicity, we write $\tilde{M}_{a|T}=p(T) M_{a|T}$. The largest value of $v$ for which a simulation is possible can then be determined by the following semidefinite program, originally introduced in Ref.~\cite{Oszmaniec2017}.
\begin{equation}
	\begin{aligned}
		v_n^{\text{crit}}=\max_{\{p\},\{\tilde{M}\}} \quad & v    \\
		\st \quad & v E_a+(1-v)\Tr(E_a)\frac{\mathbbm{1}}{d}= \sum_{T\subset S_n} \tilde{M}_{a|T},  \quad \forall a, T\\
            \quad  & \sum_{a} \tilde{M}_{a|T}=p(T)\mathbbm{1} , \quad \forall T, \\
            \quad &\tilde{M}_{a|T}=0 \quad \text{if}\quad a\notin T\\
  		\quad & \tilde{M}_{a|T}\succeq 0.
	\end{aligned}
\end{equation}
In the main text, we have evaluated this semidefinite program both when $\{E_a\}$ is the introduced qutrit SIC measurement and when $\{E_a\}$ is the tomographic reconstruction of our implemented SIC measurement.

Notably, in the case of projective measurements, no more than three outcomes can be supported in a deterministic implementation. The bound obtained from the above program, in principle valid also for nonprojective three-outcome measurements, can be saturated with rank-one projective measurements as well. This was confirmed by choosing $T=[1,2,4]$, evaluating the above semidefinite program and verifying that the optimal measurement indeed is projective. Note that because of the non-trivial phases in the Gram matrix of the SIC, not all choices of $T$ are equivalent.

\subsection*{State discrimination}
We consider also the evaluation of the success probability for discriminating the nine SIC states through a single measurement. For a given measurement $\{E_a\}$, the success probability is given by
\begin{equation}
    p_\text{suc}=\frac{1}{9}\sum_{a=1}^9 \bracket{\psi_a}{E_a}{\psi_a},
\end{equation}
where $\ket{\psi_a}$ represent the SIC states. It is well-known that qutrits cannot carry more than one trit of information \cite{Nayak1999, Tavakoli2020informationally}, which here means that $p_{\text{suc}}\leq \frac{1}{3}$ for any choice of $\{E_a\}$. It is straightforward to verify that choosing the SIC measurement saturates this optimal bound.

We are now interested in the largest value of $p_{\text{suc}}$ achievable with any stochastic combination of quantum measurements with at most $n$ outcomes. One may note that since $p_{\text{suc}}$ is linear in $E_a$, classical randomness cannot play a role for the optimal success rate. That is, the largest value of $p_{\text{suc}}$ for $n$-outcome measurements must occur for a deterministic choice of measurement. For given $n$, we can classify the measurements into $\binom{9}{n}$ different sets, each corresponding to a specific choice of which of the $9$ possible outcomes are supported by the considered measurement. The supported outcomes are listed in a set $T$. For each class, $T$, we can evaluate the largest value of $p_{\text{suc}}$ as the following semidefinite program
\begin{equation}\label{sdp1}
	\begin{aligned}
		\max_{\{E_a\}} \quad &  \frac{1}{9}\sum_{a=1}^9 \bracket{\psi_a}{E_a}{\psi_a} \\
            \st \quad  & \sum_{a} E_a = \mathbbm{1}, \\
            \quad & E_a=0 \quad \text{if}\quad a\notin T\\
  		\quad & E_a\succeq 0. 
	\end{aligned}
\end{equation}
The optimal value of $p_{\text{suc}}$ is then the largest value obtained from evaluating the $\binom{9}{n}$ semidefinite programs. In the main text, we have performed this evaluation for $n=2,\ldots,9$.

We can also permit a quantitative distrust in the preparations. Following \cite{Tavakoli2021sdi}, this is defined as a lower limit on the fidelity between the target preparation, $\ket{\psi_a}$, and the potentially mixed and misaligned lab preparation $\rho_a$. The distrust bound corresponds to
\begin{equation}
    F(\psi_a,\rho_a)=\bracket{\psi_a}{\rho_a}{\psi_a}\geq 1-\epsilon_a,
\end{equation}
where $\epsilon_a$ is the distrust parameter. We have measured $\epsilon_a$ in the lab and obtained, 
\begin{equation}
\epsilon=(0.0084, 0.002, 0.011, 0.008, 0.0061, 0.0115, 0.0058, 0.0089, 0.004).
\end{equation}
Some variation is observed over the different ports. Then, to estimate the worst-case correction to the maximal value of $p_{\text{suc}}$ under $n$-outcome measurements, as a consequence of this level of distrust in the preparations, we have employed semidefinite programs in seesaw \cite{tavakoli2023semidefinite}. Specifically, we select a random set of states, and evaluate a program analogous to \eqref{sdp1} (replacing $\ket{\psi_a}$ with the randomly sampled states). Using the returned measurement, we evaluate the semidefinite program
\begin{equation}\label{sdp2}
	\begin{aligned}
		\max_{\{\rho_a\}} \quad &  \frac{1}{9}\sum_{a=1}^9 \Tr(\rho_a E_a) \\
            \st \quad  & \Tr(\rho_a)=1, \quad \forall a, \\
            \quad & \bracket{\psi_a}{\rho_a}{\psi_a}\geq 1-\epsilon_a \quad \forall  a\\
  		\quad & \rho_a\succeq 0. 
	\end{aligned}
\end{equation}
The optimal states are then used to run the semidefinite program over the measurements etc. until sufficient convergence is seen. Naturally, this is an interior point search and does not guarantee a global optimum. We have therefore repeated this procedure 50 times for every choice of  $T$ and selected the largest result. These results are presented in the main text.

\section{Random number generation}
We generate random numbers from the SIC measurement in a measurement-device-independent scenario. That is, the security model makes no assumptions on the measurement but assumes that it is probed with the nine states $\rho_x$. The randomness is evaluated from the distribution obtained from one of the nine states, denoted $x^*\in\{1,\ldots,9\}$. The randomness contained in the distribution $p(a|x^*)$ is defined as the negative logarithm of the largest probability with which an eavesdropper with quantum control of the measurement device can guess its output. Thus, the randomness is $R=-\log_2(P_g)$, where the guessing probability is
\begin{equation}
    P_g=\max \sum_{a=1}^9 p(a,e=a|x^*),
\end{equation}
where $p(a,e|x^*)$ is the joint distribution of the device and the eavesdropper.

In general, the eavesdropper can share an entangled state, $\sigma^{AE}$, with the device and control the device so that it performs a measurement $\{M_a\}$ on the joint system $PA$, where the probe system $P$ corresponds to the incoming state $\rho_x$. She can also measure her own system, $E$, with some POVM $\{N_e\}$. Following Ref.~\cite{Skrzypczyk2017}, the resulting correlations can then be written as
\begin{equation}
    p(a,e|x)=\Tr\left( (M_a\otimes N_e)(\rho_x^P\otimes \sigma^{AE})\right)=\Tr\left( \rho_x \tilde{M}_{a,e}\right),
\end{equation}
where we defined $\tilde{M}_{a,e}=\Tr_{AE}\left((M_a\otimes N_e)(\mathbbm{1}^P\otimes \sigma^{AE})\right)$. It is then possible to formulate the guessing probability in terms of a semidefinite program
\begin{equation}\label{sdp3}
	\begin{aligned}
		P_g=\max_{\{p\},\{\tilde{M}\}} \quad &  \frac{1}{9}\sum_{a=1}^9 \Tr\left( \rho_{x^*} \tilde{M}_{a,a}\right)\\      
  \st \quad  &  p(a|x)=\sum_e \Tr\left( \rho_{x^*} \tilde{M}_{a,e}\right), \quad \forall a,x, \\
            \quad  & \sum_{a}  \tilde{M}_{a,e}=p(e)\mathbbm{1}, \quad \forall e, \\
            \quad & \sum_{e}p(e)=1\\
  		\quad & \tilde{M}_{a,e}\succeq 0. 
	\end{aligned}
\end{equation}
The first condition corresponds to recovering the experimentally observed probabilities as marginals over the eavesdropper. The second constraint follows from the properties of $\{\tilde{M}_{a,e}\}$ and may be understood as no-signaling. The third property is normalisation of the effective measurement. See Ref.~\cite{Skrzypczyk2017} for further details. 

\begin{figure}
    \centering
    \includegraphics[width=0.7\columnwidth]{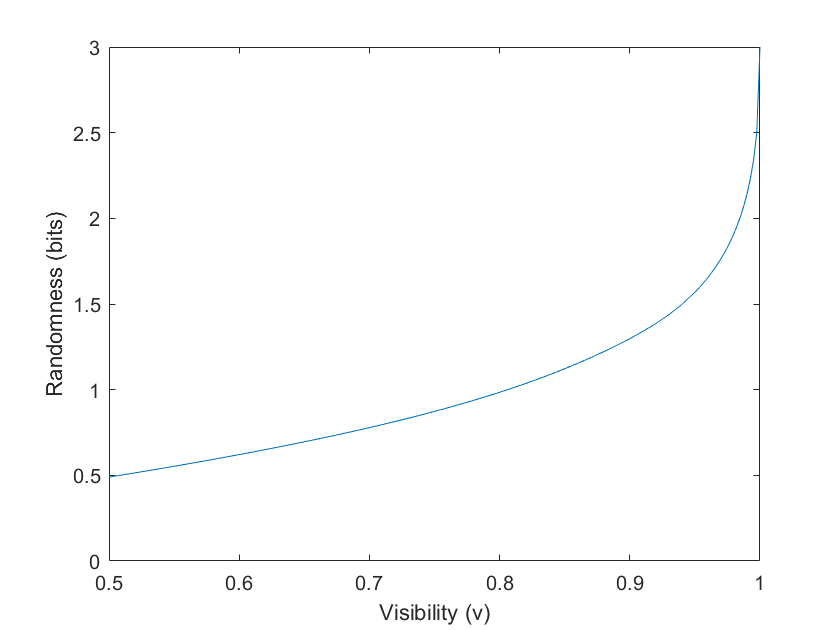}
    \caption{\textbf{Randomness versus visibility.} Measurement-device-independent randomness versus visibility of SIC measurement using the quantum state exclusion protocol.}
    \label{supfig4}
\end{figure}

We first apply the above semidefinite program to our protocol for a noisy SIC measurement of the form
\begin{equation}
    E_a^{(v)}=v E_a^{\text{SIC}}+(1-v)\frac{\mathbbm{1}}{9}.
\end{equation}
We choose to extract randomness from the first input, $x^*=1$.
The dependence of the randomness on the parameter $v$ is displayed in Fig.~\ref{supfig4}. Indeed, $v=1$ gives $\log_2(8)=3$ bits of randomness. However, as expected, the yield falls quickly as $v$ decreases from its ideal value. Already at $v\approx 0.9525$, one reaches the projective randomness limit of $R=\log_2(3)$ bits. 

Firstly, if we assume that the noise is isotropic in our device, it was shown in the simulation section that our device's visibility must exceed the six-outcome simulation limit of $v=0.9571$, which is above the required threshold for beating $\log_2(3)$ bits. Secondly, we may use the tomographic reconstruction of the lab measurement as input to the semidefinite program. This leads to $R=1.7973$ bits per round. Thirdly, another approach, which avoids the drawbacks of tomographic reconstruction, is to use the relative frequencies obtained in the quantum state exclusion protocol described in the main text for random number generation. However, as the relative frequencies obtained in the experiment are not exact probabilities, we replace the first constraint in the semidefinite program with a tolerance region, corresponding to a radius $r_{a,x}\geq 0$, 
\begin{equation}
    p(a|x)-r_{a,x}\leq  \sum_e \Tr\left( \rho_{x^*} \tilde{M}_{a,e}\right) \leq p(a|x)+r_{a,x}.
\end{equation}
It turns out that we can choose a large tolerance region and still outperform the projective randomness limit. For instance, choosing each $r_{a,x}$ to be ten percent of the mean, every choice of $x^*$ leads to significantly more randomness than $\log_2(3)$ bits. The best choice is $x^*=7$ which gives $R=1.7198$ bits. Fourthly, we can also choose the tolerance regions based on Poissonian statistics. Again choosing a large tolerance region, we may choose $r_{a,x}$ as, for instance, 12 standard deviations of the mean relative frequency when $a\neq x$. When $a=x$  the means are much smaller, making 12 standard deviations comparable to the mean. We therefore choose a still significant five standard deviations for $r_{a,a}$. This returns $R=1.7199$ bits from the setting $x^*=7$. In conclusion, while there are many ways of analysing the randomness in the experiment, they all lead to bit rates that clearly exceed the fundamental limitations of projective measurements.

\section{Beating real-valued four-dimensional quantum systems}

\begin{figure}[h]
\centering
\includegraphics[width=15.0cm]{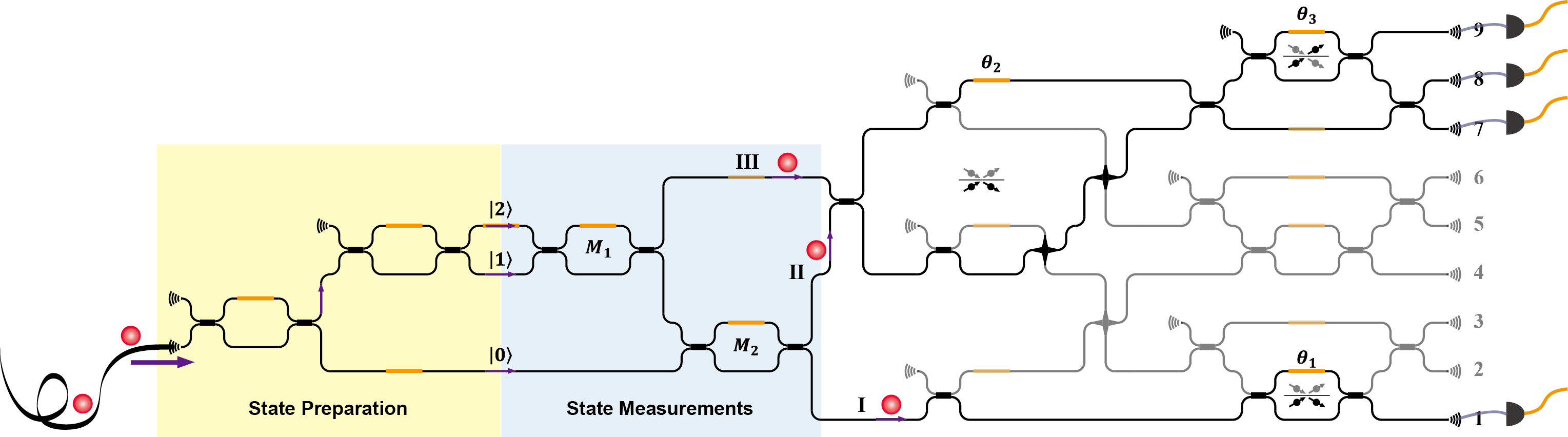}
\caption{\textbf{Experiment setting for MUB test.} The black line routes the expected measurements to the output port. Phases $\theta_1$ and $\theta_2$ are fixed to maintain the interferometer fully reflected, and phase $\theta_3$ is fixed to maintain the interferometer fully transmitted (shown in insets).} 
\label{supfig5}
\end{figure}

Four mutually unbiased bases (MUBs) in three-dimensional Hilbert space are given by
\begin{align}\label{MUBs}
&\begin{Bmatrix}
1 & 0 & 0\\
0 & 1 & 0\\
0 & 0 & 1
\end{Bmatrix}
& \begin{Bmatrix}
1 & 1 & 1 \\
1 & \omega & \omega^2 \\
1 & \omega^2 & \omega
\end{Bmatrix}
&& \begin{Bmatrix}
1 & \omega & \omega\\
\omega & 1 & \omega\\
\omega & \omega & 1
\end{Bmatrix}
&&\begin{Bmatrix}
1 & \omega^2 & \omega^2\\
\omega^2 & 1 & \omega^2\\
\omega^2 & \omega^2 & 1
\end{Bmatrix},
\end{align}
where the columns are the basis states, associated with outcome $a=0,1,2$. The four bases can be represented by two bits, $y_0,y_1\in\{0,1\}$, or alternatively $y=2y_0+y_1+1$. 

Recall now the SIC set, 
\begin{equation}
\begin{Bmatrix}
& (0,1,-1) & (0,1,-\omega) & (0,1,-\omega^2)\\
& (-1,0,1) & (-\omega,0,1) & (-\omega^2,0,1)\\
& (1,-1,0) & (1,-\omega,0) & (1,-\omega^2,0)
\end{Bmatrix}.
\end{equation}
It can be labeled by a pair of trits, $x_0,x_1\in\{0,1,2\}$, where $x_0$ labels the row and $x_1$ labels the column above. We may alternatively write $x=3x_0+x_1+1$. 

Compare now the SIC states with the four bases. Some inspection shows that any state selected from the SIC set can be measured in any of the four bases and yield, up to permutation, the probability distribution $(0,1/2,1/2)$ over the three possible outcomes. In order to characterise the distribution, we must identify the value of the outcome $a$ to which the zero probability is associated for each $(x,y)$. In Ref.~\cite{Huang2021}, the following rule is identified for the zero-probability event,
\begin{equation}
    \begin{cases}
        a=x_{y_1} & \text{if } y_0=0\\
        a=x_{0}+(-1)^{y_1}x_{1} \mod{3} & \text{if } y_0=1
    \end{cases}. 
\end{equation}

In a communication game in which one party sends one of nine arbitrary states and another selects one of four arbitrary measurements, we define the losing condition as satisfying the above relation, i.e.~outputting the 'wrong' value of $a$. The average failure rate is given by
\begin{equation}
    S=\frac{1}{36}\sum_{x=1}^9 \sum_{y=1}^4 p(\text{condition satisfied}|x,y).
\end{equation}
In the qutrit protocol based on sending SIC states and measuring the four mutually unbiased bases, we achieve a perfect score of $S=0$. 

Using a slightly different, but ultimately equivalent, scoring function, it was evidenced in \cite{Pauwels2022} that there exists no strategy based on states in $\mathbf{R}^4$ that can achieve the perfect score. This may appear counter-intuitive, as higher-dimensional quantum systems have a higher information capacity and typically outperform their lower-dimensional counterparts. This is, however, not true in general. Following the same numerical methods as in \cite{Pauwels2022}, but now for the function $S$, we find that such protocols cannot perform better than $S\approx 8.0\times 10^{-3}$. Thus, an experiment that achieves a smaller score falsifies any quantum model based on four-dimensional real-valued systems. This is particularly interesting as the correlations achievable by such systems have a natural operational meaning in terms of the correlations achievable in entanglement-assisted qubit communication, in any protocol in the spirit of dense coding, where parties share an EPR pair and perform unitary encodings \cite{Pauwels2022}.

\begin{figure}[ht]
\centering
\includegraphics[width=15.0cm]{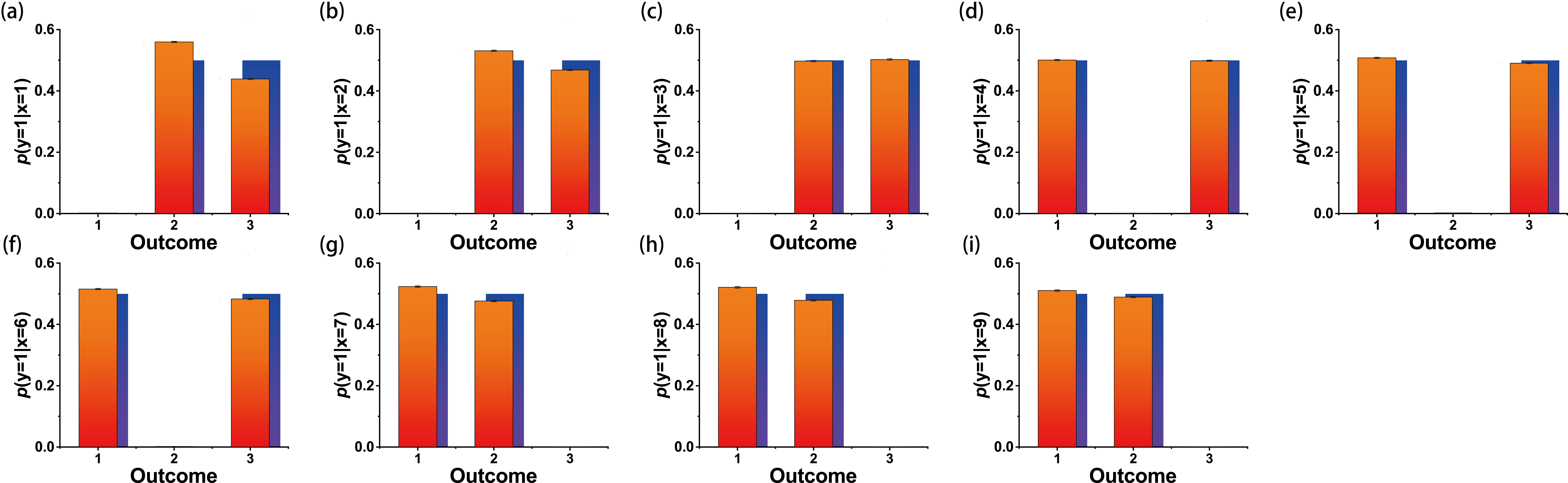}
\caption{\textbf{Probability distribution of the first MUB measurement.} (a)-(i) represents the probability of each output when the input state is $|\psi_1\rangle-|\psi_9\rangle$. The horizontal axis represents the outcome and the vertical axis represents the probability of each outcome. The blue column represents the theoretical value. The red column represents the experimental value.}
\label{supfig6}
\end{figure}

\begin{figure}[ht]
\centering
\includegraphics[width=15.0cm]{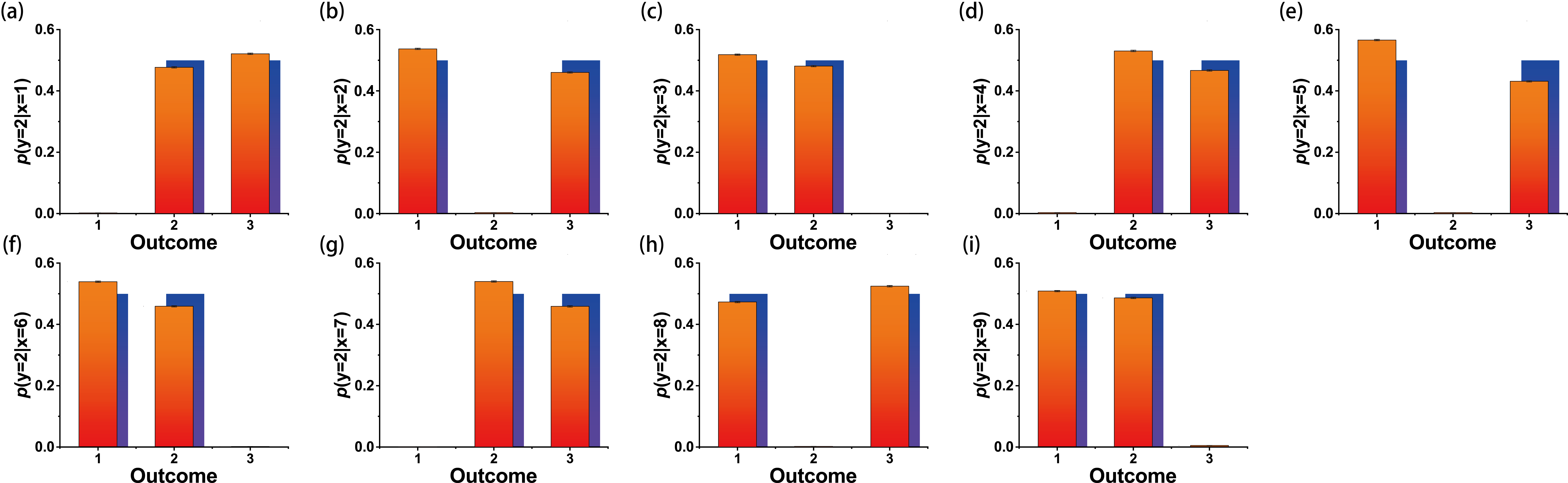}
\caption{\textbf{Probability distribution of the second MUB measurement.} (a)-(i) represents the probability of each output when the input state is $|\psi_1\rangle-|\psi_9\rangle$. The horizontal axis represents the outcome and the vertical axis represents the probability of each outcome. The blue column represents the theoretical value. The red column represents the experimental value.}
\label{supfig7}
\end{figure}

\begin{figure}[ht]
\centering
\includegraphics[width=15.0cm]{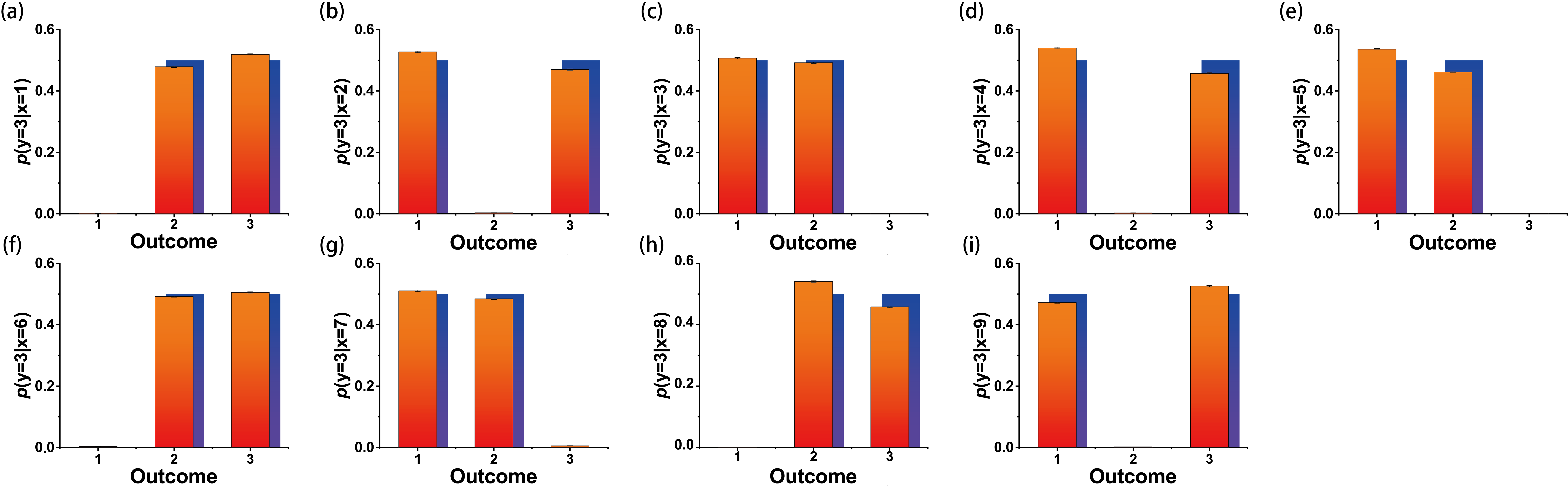}
\caption{\textbf{Probability distribution of the third MUB measurement.} (a)-(i) represents the probability of each output when the input state is $|\psi_1\rangle-|\psi_9\rangle$. The horizontal axis represents the outcome and the vertical axis represents the probability of each outcome. The blue column represents the theoretical value. The red column represents the experimental value.}
\label{supfig8}
\end{figure}

\begin{figure}[ht]
\centering
\includegraphics[width=13.0cm]{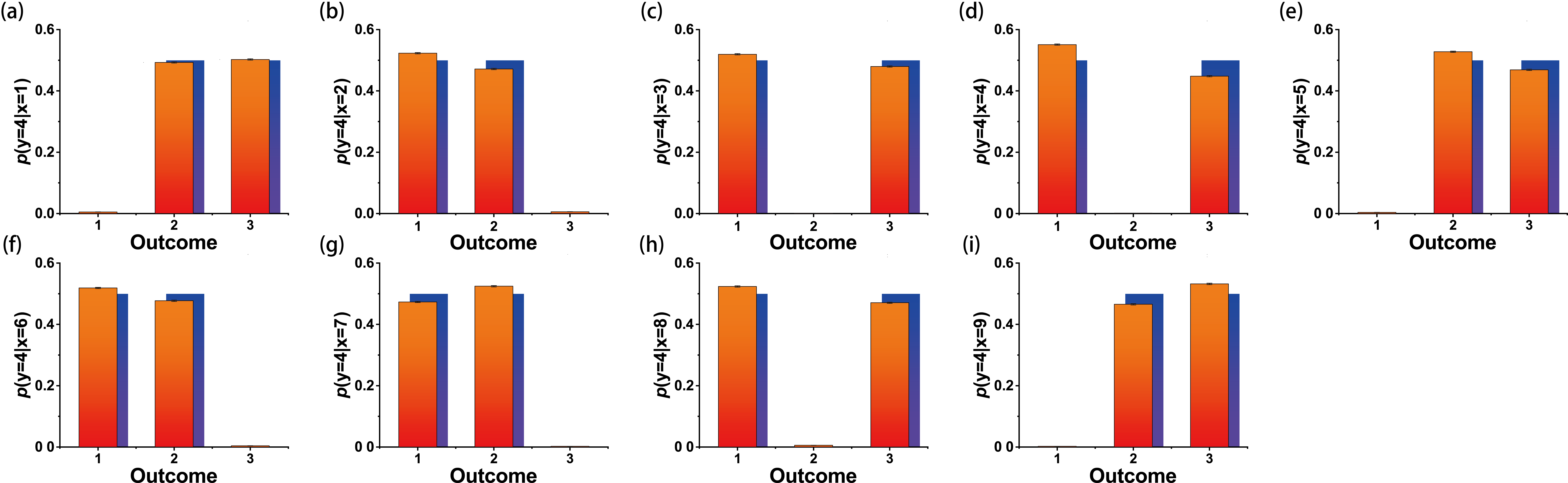}
\caption{\textbf{Probability distribution of the fourth MUB measurement.} (a)-(i) represents the probability of each output when the input state is $|\psi_1\rangle-|\psi_9\rangle$. The horizontal axis represents the outcome and the vertical axis represents the probability of each outcome. The blue column represents the theoretical value. The red column represents the experimental value.}
\label{supfig9}
\end{figure}

The experiment setting for the MUB test is shown in Fig.~\ref{supfig5}, and the \textit{State Measurements} region can realize any MUB measurements and get outcomes in three different paths (Paths I, II, and III). The follow-up circuit is reconfigured to route the results to different output ports (Path I $\rightarrow$ Port 1, Path II $\rightarrow$ Ports 7 and 8, and Path III $\rightarrow$ Port 9). Probability distributions for MUB test are shown in ~\cref{supfig6,supfig7,supfig8,supfig9}, and our experimental result achieves $S\approx 2.4\times 10^{-3}$. Using Hoeffding's inequality, it falsifies the real-valued four-dimensional model with a vanishingly small $p$-value. This is a consequence of our sum-total of roughly $1.6$ million counts. The full set of relative frequencies in this experiment are illustrated below. In particular, we see that the worst-case failure probability, selected over all $(x,y)$, is no more than $0.0018$.

\section{Randomised Matching task}
We study a task in which Alice and Bob independently, privately and uniformly at random select one of $n$ possible symbols, $x,y\in\{1,\ldots,n\}$, repectively. Alice 
sends a message from an alphabet of size $d=n-1$  to Bob, upon which he must answer the question whether their selections were identical or not. The average success rate of the random matching question is
\begin{equation}
	T= \frac{1}{n^2} \sum_{x,y=1}^n p(a=\delta_{x,y}|x,y).
\end{equation}
To calculate the maximal value of $R$, we use the completeness of the measurement to obtain
\begin{align}
	T &= \frac{1}{n^2} \qty{\sum_{x\ne y} \tr \qty(\rho_x M_{0|y}) +\sum_{x= y} \qty(\rho_x M_{1|y})  } \\
	&= \frac{1}{n^2} \qty{ \sum_y \tr \qty( M_{1|y} \sum_x (-1)^{\delta_{xy}+1} \rho_x) + n^2-n } \\
	&= \frac{n-1}{n} + \frac{1}{n^2}\sum_y \tr \qty(M_{1|y} \mathcal{O}_y) \, , \label{eq:SOmax}
\end{align}
where we have defined 
\begin{equation}
	\mathcal{O}_y \equiv \rho_y - \sum_{x \ne y} \rho_x \, . \label{eq:O}
\end{equation}
Since $T$ is linear in the states, they are optimally taken to be pure. Consequently, each $\mathcal{O}_y$ has only one positive eigenvalue. We optimally choose $M_{1|y}$ as the projector onto the corresponding eigenvector, leading to
\begin{equation}
	T= \frac{n-1}{n} + \frac{1}{n^2}\sum_y \lambda_{\rm max}( \mathcal{O}_y) \, , \label{eq:SLambdaMax}
\end{equation}
where $\lambda_{\rm max}( \mathcal{O}_y)$ is the maximal eigenvalue of $\mathcal{O}_y$. 

Consider now a classical model, in which all states are diagonal in the same basis. Without loss of generality, we can write such states as $	\rho_x = \sum_{m=1}^d p(m|x) \ketbra{m}{m}$,  
where $p(m|x)$ is the probability for Alice to send the message $m$ when she gets the input $x$. Due to linearity, the optimal choice is deterministic, i.e.~$p(m|x)=\delta_{f(x),m}$ for some function $f:\{1,\ldots,n\}\rightarrow \{1,\ldots,n-1\}$. This leads to $\lambda_{\text{max}}(\mathcal{O}_y)=\max_m \delta_{f(y),m}-\sum_{x\neq y} \delta_{f(x),m}$. Clearly, we need only to distinguish $f(y)$ from all other values $f(x)$ (for any $x\neq y$). An optimal choice is  $f(x)=x$  for every $x=1,\ldots,n-1$, and then making any choice for $f(x=n)$.  This leads to
\begin{align}
	T_{\text{C}} = 1-\frac{2}{n^2},
\end{align}

Now, consider instead a quantum model based on equiangular projections. Specifically, the outcome $a=1$ for any selection of $y$ is associated to a $d$-dimensional rank-one projector $\ket{\phi_y}$ with the property that 
\begin{equation}
    |\braket{\phi_y}{\phi_{y'}}|^2=\frac{1}{d^2},
\end{equation}
for every $y\neq y'$. The constant is known to be the smallest possible such that all overlaps are identical. This property is known to imply that $\sum_{y=1}^n \ketbra{\psi_y}{\psi_y} = \frac{n}{d} \mathbbm{1}$. Such states can be constructed in any dimension by removing a row from the Fourier transform matrix and re-normalising the columns. Inserted in our expression for $\mathcal{O}_y$, we obtain the simple expression
\begin{equation}
    \mathcal{O}_y=\ketbra{\phi_y} -\sum_{x\neq y} \ketbra{\phi_x}
    =2\ketbra{\phi_y} -\sum_{x} \ketbra{\phi_x}=2\ketbra{\phi_y}-\frac{n}{d}\mathbbm{1}.
\end{equation}
Clearly, the largest eigenvalue is $\lambda_{\text{max}}(\mathcal{O}_y)=2-\frac{n}{d}$. This leads to the quantum value
\begin{align}
	T_{\text{Q}}= 1-\frac{1}{n(n-1)},
\end{align} 
which is strictly larger than the classical limit. 

\begin{figure}[ht]
\centering
\includegraphics[width=10.0cm]{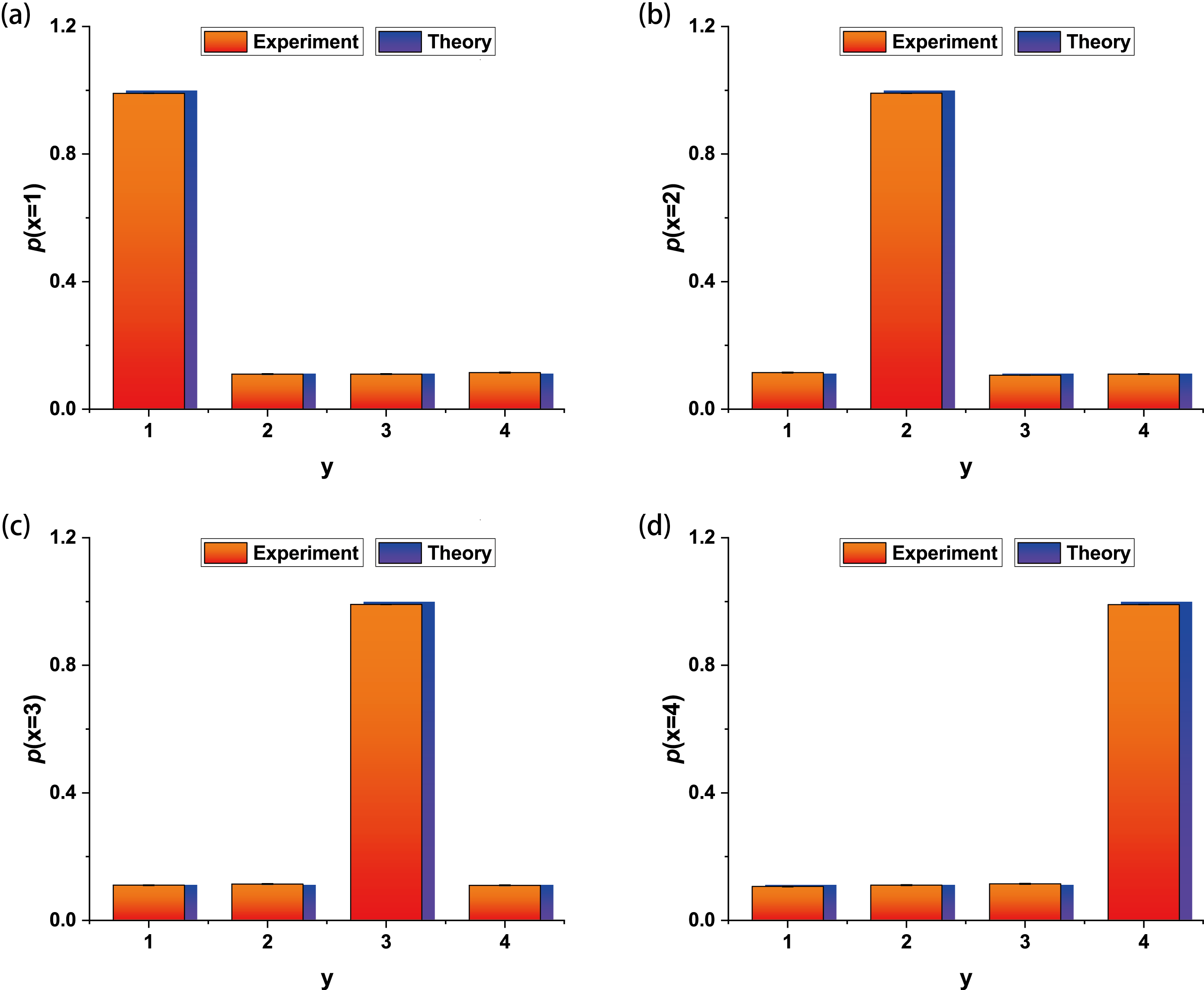}
\caption{\textbf{Probability distribution for randomised matching protocol.} (a)-(d) represents the probability of observing outcome $a=1$ (matching)  when the input state is for each of the four input states.  The blue column represents the theoretical value. The red column represents the experimental value.}
\label{supfig10}
\end{figure}

The probability distribution estimated from our experimental implementation is shown in Fig.~\ref{supfig10}.

\begin{refcontext}[sorting=none]
\printbibliography
\end{refcontext}